\documentstyle[12pt]{article}
\newcommand{\beq}{\begin{eqnarray}}
\newcommand{\eeq}{\end{eqnarray}}

\newcommand{\ep}{{\epsilon}}

\newcommand{\f}{\frac}

\newcommand{\ra}{\rightarrow}

\begin{document}

\topmargin 0pt

\oddsidemargin -3.5mm

\headheight 0pt

\topskip 0mm
\addtolength{\baselineskip}{0.20\baselineskip}
\begin{flushright}
\end{flushright}
\begin{flushright}
hep-th/0306152v4
\end{flushright}
\vspace{0.5cm}
\begin{center}
    {\large \bf  Statistical Mechanics of  Three-dimensional Kerr-de Sitter
Space }
\end{center}
\vspace{0.5cm}
\begin{center}
 Mu-In Park\footnote{~Electronic address: muinpark@yahoo.com} \\
 \vspace{0.3cm}
{\it   Research Institute of Physics and Chemistry,\\ Chonbuk
National University, Chonju 561-756, Korea}
\end{center}
\vspace{0.5cm}
\begin{center}
    {\bf ABSTRACT}
\end{center}
The statistical entropy computation of the (2+1)-dimensional Kerr-de
Sitter space in the context of the {\it classical} Virasoro algebra
for an asymptotic isometry group has been a mystery since first, the
degeneracy of the states has the right value only at the infinite
boundary which is casually disconnected from our universe, second,
the analyses were based on the unproven Cardy's formula for complex
central charge and conformal weight. In this paper, I consider the
entropy in Carlip's ``would-be gauge" degrees of freedom approach
instead. I find that it agrees with the Bekenstein-Hawking entropy
but there are no the above problems. Implications to the dS/CFT are
noted. \vspace{3cm} \vspace{2.5cm}
\begin{flushleft}

06 Feb. 2009 \\
\end{flushleft}

\newpage

After recent cosmological observations that suggest the existence of
a positive cosmological constant $\Lambda$ there have been extensive
studies on the de-Sitter space. Especially, in the context of dS/CFT
proposal \cite{Park:98,Park:99NP,Hull:98,Stro:01} \footnote{~Ref.
\cite{Park:98} provides the first explicit example of the dS/CFT in
the three-dimensional Kerr-de Sitter space. In Ref.
\cite{Park:99NP}, as the direct consequence of Ref. \cite{Park:98},
``dS/CFT'' for the three-dimensional case was noted explicitly for
the first time: See, for example, below Eq. (47).}, there have been
increasing interests \cite{Klem:02} in statistical computation of
the Bekenstein-Hawking (BH) entropy for the three-dimensional
Kerr-de Sitter $(KdS_3)$ space \cite{Gibb:77}
\begin{eqnarray}
\label{BH}
 S=\f{2 \pi r_{+}}{4 \hbar G},
\end{eqnarray}
from the ``classical'' Virasoro algebra, associated with an
asymptotic isometry group at the infinite boundary {\it \`a~la}
Strominger and Birmingham \cite{Stro:98}. Here, $r_+$ is the radius
of the cosmological horizon. But, there have been two problems in
those analyses. The first is a conceptual problem that the entropy
was computed at the infinite boundary which is casually disconnected
from the observer who is sitting in our universe, the interior
($r<r_+$) of the cosmological horizon. The second is that the
analyses were based on a plausible but unproven assumption about the
extended Cardy's formula \cite{Card:86} for the {\it complex}
central charge and conformal weight. (See \cite{Bous:01} and
\cite{Park:0705} for recent strugglings with this problem.)

In this paper I consider the entropy in the context of Carlip's
``would-be gauge" degrees of freedom approach \cite{Carl:95} instead
and show that the above two problems are resolved. The use is made
of $SL(2, {\bf C})$ WZW theory {\it on the horizon} and its
associated {\it well-defined} Virasoro operator. The obtained
entropy agrees with the BH entropy (\ref{BH}) but there are no the
problems above. Implications to the dS/CFT are noted also.

The (2+1)-dimensional gravity for a cosmological constant $\Lambda
=1/{\ell}^{2}$ is described by the action
\begin{eqnarray}
\label{gravity} I_{g}=\frac{1}{16 \pi G} \int d^3 x \sqrt{-g} \left(
R -\frac{2}{ {\ell}^{2}} \right)+I_m,
\end{eqnarray}
where $I_m$ is the matter action [the details are not important in
this paper] and I have omitted the surface terms in the pure gravity
part. This theory has a constant curvature $R=2 {\ell}^{-2}$,
outside the matters.

For the positive cosmological constant, {\it i.e.}, real valued
${\ell}$, (\ref{gravity}) has the Kerr-type vacuum solution, called
three-dimensional Kerr-de Sitter ($KdS_3$),
(from now on,
I take $G=1/8$ unless otherwise stated)
\begin{eqnarray}
\label{kds3} ds^2 =-\left(M-\frac{r^2}{{\ell}^2} +\frac{J^2}{4 r^2}
\right) dt^2 +\left( M-\frac{r^2}{{\ell}^2} +\frac{J^2}{4
r^2}\right)^{-1} dr^2 +r^2 \left(-\f{J}{2 r^2}  dt +d
{\varphi}\right)^2
\end{eqnarray}
with the mass and angular momentum parameters $M$ and $J$,
respectively
\cite{Park:98,Stro:01,Bous:01,Park:0705,Dese:84,Myun:01}; $\varphi$
has period $2 \pi$. This solution has a cosmological horizon at
\begin{eqnarray}
r_{+} =\frac{{\ell}}{\sqrt{2}} \sqrt{M+\sqrt{M^2
+\frac{J^2}{{\ell}^2}}}
\end{eqnarray}
with no bound on $M$ and $J$. By introducing $r_{(-)} \equiv
({{\ell}}/{\sqrt{2}})\sqrt{-M+\sqrt{M^2 +{J^2}/{{\ell}^2}}}$, the
metric (\ref{kds3}) can be conveniently written in the proper radial
coordinates as
\begin{eqnarray}
ds^2 =\mbox{sinh}^2 \rho \left( \frac{r_+ dt}{{\ell}} -r_{(-)} d
{\varphi} \right) ^2 -{\ell}^2 d \rho^2 +\mbox{cosh}^2 \rho \left(
\frac{r_{(-)} dt}{{\ell}} +r_+ d \varphi \right )^2
\end{eqnarray}
with
\begin{eqnarray}
M=\frac{r_+^2 -r_{(-)}^2}{{\ell}^2}, ~ J=\frac{2 r_+ r_{(-)}}{{\ell}}, \nonumber \\
\nonumber \\
r^2=r^2_+ \mbox{cosh}^2 \rho + r^2_{(-)} \mbox{sinh}^2 \rho.
\end{eqnarray}
In these coordinates, the cosmological event horizon is at $\rho
=0$, and the exterior ($r>r_+$)/interior ($r<r_+$) of the horizon is
represented by real/imaginary value $\rho$. The interior and
exterior regions are casually disconnected and the horizon acquires
the thermodynamic entropy (\ref{BH}). I seek a statistical
derivation of (\ref{BH}) in the context of the (2+1)-dimensional
quantum gravity with {\it would-be gauge} degrees of freedom ``on
the horizon'', {\it \`a la} Carlip \cite{Carl:95}.

The quantization of (\ref{gravity}) is best achieved by recasting
the theory as a $SL(2, {\bf C})$ Chern-Simons gauge theory
\cite{Achu:86,Witt:91}. The action for this theory is, up to
surface terms,
\begin{eqnarray}
\label{CS}
 I_g [A]=\frac{is}{4 \pi} \int_{D_2 \times R} d^3 x ~
\epsilon^{\mu \nu \rho} \left<A_{\mu} \partial _{\nu} A_{\rho}
+\frac{2}{3} A_{\mu} A_{\nu} A_{\rho} \right> -\frac{is}{4 \pi}
\int _{D_2 \times R} d^3 x ~\epsilon^{\mu \nu \rho} \left<
\bar{A}_{\mu}
\partial _{\nu} \bar{A}_{\rho} +\frac{2}{3} \bar{A}_{\mu}
 \bar{A}_{\nu} \bar{A}_{\rho} \right>
\end{eqnarray}
on the manifold $\Sigma=D_2 \times R$, where $D_2$ is a
2-dimensional disc of the space at a constant time $t$, with its
boundary $\partial D_2$ at the cosmological horizon $r_+$;
${A}_{\mu}$ is an $SL(2, {\bf C})$ gauge field and $\bar{A}_{\mu}$
is its complex conjugate; $ \left< \cdots \right>$ denotes the
trace. (\ref{CS}) is equivalent to (\ref{gravity}) with the
identification \footnote{~ I use the same conventions as Ref.
\cite{Park:98}: $ t_0=\frac{1}{2} \left(\begin{array}{cc} 0 & -1
\\ 1 & 0
\end{array} \right) ,~ t_1=\frac{1}{2} \left(\begin{array}{cc} 1 &
0 \\ 0 & -1
\end{array} \right),~ t_2 =\frac{1}{2} \left( \begin{array}{cc} 0
& 1 \\ 1 & 0 \end{array}
 \right)$ such that
$[t_a, t_b]={\epsilon_{ab}}^c t_c$ and $\left< t_a t_b \right>
=\frac{1}{2} \eta_{ab}$, where $\epsilon_{012}=1$ and
$\eta_{ab}=diag(-1,1,1)$. The metric $\eta_{ab}$ is used to raise
and lower indices.}
\begin{eqnarray}
A^a_{\mu}=\omega^a_{\mu}+\frac{e^a_{\mu}}{i {\ell}},~
\bar{A}^a_{\mu}=\omega^a_{\mu}-\frac{e^a_{\mu}}{i {\ell}}
~~(a=0,1,2)
\end{eqnarray}
with
\begin{eqnarray}
\label{s}
 s=2{\ell}.
\end{eqnarray}
Here, $e^a=e^a_{\mu}d x^{\mu},~ \omega^a=\frac{1}{2}
\epsilon^{abc} \omega_{\mu b c} d x^{\mu}$ are the triads and the
$SL(2, {\bf R})$ spin connections, respectively. It is easily
checked that the $KdS_3$ solution (\ref{kds3}) is represented by
\begin{eqnarray}
{\bf A}^0& =&-\frac{r_+ +i r_{(-)}}{{\ell}} \left( \frac{dt}{{\ell}}
+i d {\varphi}
\right) \sinh \rho, \nonumber \\
{\bf A}^1 &=& d \rho, \nonumber \\
{\bf A}^2 &=&-\frac{r_+ +ir _{(-)}}{{\ell}} \left( \frac{dt}{{\ell}}
+i d \varphi \right) \cosh \rho,
\end{eqnarray}
where the superscript indices denote the group indices $a=0,1,2$.
From this, the polar components are obtained as
\begin{eqnarray}
\label{A}
 A_{\rho}=t_1,~A_{\varphi}=-i{z} ~({U}^{-1} t_2 {U} ),~
A_t =i A_{\varphi},
\end{eqnarray}
where
\begin{eqnarray}
 {z}=(r_+ +i r_{(-)})/{\ell},~~
{U} = \left(
\begin{array}{cc}
e^{\rho /2} &0 \\
0 & e^{-\rho/2}
\end{array} \right).
\end{eqnarray}
This insures that there is asymptotic isometries as $\rho
\rightarrow \infty$ and its associated ( two-copies of )
``classical'' Virasoro algebra \cite{Park:98}. The obtained
statistical entropy coincides with (\ref{BH}), which means that the
appropriate CFT which gives the correct entropy of a horizon would
live {\it at the infinite boundary}, as coincides exactly with
Strominger's dS/CFT proposal \cite{Stro:01}. But, in the following,
I will show that (\ref{A}) {\it also} allows to acquire the BH
entropy (\ref{BH}) from a quantum mechanical analysis of $SL(2, {\bf
C})$ WZW conformal field theory {\it on the horizon} $r_+$ which is
located at the space boundary $\partial D_2$, by a direct adaptation
of Carlip's analysis \cite{Carl:95}.

To this end, I first note that the gauge connection (\ref{A})
satisfies the boundary conditions for a black hole horizon $r_+$,
even if $r_{(-)}$ is an arbitrary real parameter, 
$\tilde{r}_{(-)}$ \cite{Bana:96}. This means that we have to  sum
over its possible values to count macroscopically {\it
in}-distinguishable states \cite{Carl:95}. In the following I will
find the value of $\tilde{r}_{(-)}$ which gives the dominant
contributions to the entropy.

Imposing the Gauss' law constraint $F^a_{ij}=0$ reduces (\ref{CS})
to a $SL(2, {\bf C})$ chiral WZW action
\cite{Park:98,Moor:89,OhPark:98,Fjel:99}, up to the surface terms
which vanish on the horizon $r_+$ \cite{Bana:96},
\begin{eqnarray}
I_g [A]&=&I_{CWZW} [A] -I_{CWZW} [\bar{A}], \\
I_{CWZW}[A]&=&-\f{is}{4 \pi} \int_{D_2} \epsilon_{ij} \left< A_i
A_j A_0 \right> +\f{is}{4 \pi} \oint_{\partial D_2} d \varphi
\left< A_{\varphi} A_0 \right> \nonumber
\end{eqnarray}
with
\begin{eqnarray}
A_{\mu} =g^{-1} \partial_{\mu} g .
\end{eqnarray}
The Poisson bracket for $A_{\varphi}^a$, {\it which lives on
$\partial D_2=r_+$}, is computed \cite{Witt:84} as
\begin{eqnarray}
\{A_{\varphi}^a(\varphi), A_{\varphi}^b (\varphi ') \} & =&\frac{4
\pi} {is}
 {\epsilon^{ab}}_{c}~ A_{\varphi}^c
(\varphi) \delta(\varphi -\varphi ') +\frac{ 4 \pi} {is} \eta^{ab}
\partial_{\varphi} \delta (\varphi -\varphi ')
\end{eqnarray}
which is the $SL(2, {\bf C})$ Kac-Moody algebra in the density form
\cite{Park:98,Moor:89,OhPark:98}. The current operators $J_n^a,
\bar{J}_n^a$ which are defined by
\begin{eqnarray}
\label{J} A_{\varphi}^a =\f{2 }{is \hbar}
\sum_{n=-\infty}^{\infty} J^a_n e^{in \varphi},
~~~\bar{A_{\varphi}}^a =-\f{2 }{is \hbar}
\sum_{n=-\infty}^{\infty} \bar{J}_n^a e^{in \varphi}
\end{eqnarray}
satisfy $SL(2, {\bf C})$ operator current algebra
\begin{eqnarray}
\label{JJ}
 &&[ J_m^a, J_n^b ] = i {\epsilon^{ab}}_c J_{m+n}^c +k m
\eta^{ab}
\delta_{m+n,0} , \nonumber \\
&&[ J_m^a, \bar{J}_n^b] = 0, \\
&&[ \bar{J}_m^a, \bar{J}_n^b] = i {\ep^{ab}}_c \bar{J}_{m+n}^c -k
m \eta^{ab} \delta_{m+n,0} \nonumber
\end{eqnarray}
with the imaginary level
\begin{eqnarray}
k=-\f{i s}{2 \hbar}.
\end{eqnarray}
The Virasoro operators $L_0,~\bar{L}_0$ for the action  (\ref{CS})
are then given by the Sugawara construction \cite{Fran:97} as
\begin{eqnarray}
\label{L0}
 L_0 =\f{1}{2k +Q} \sum^{\infty}_{n=-\infty} \left<:J_{-n}
J_n :\right>',~~ \bar{L}_0 =\f{1}{-2k +Q} \sum^{\infty}_{n=-\infty}
\left<:\bar{J}_{-n} \bar{J}_n :\right>',
\end{eqnarray}
which satisfy
\begin{eqnarray}
[L_0, J^a_n ]=-n J^a_n,~~[\bar{L}_0, \bar{J}^a_n ]=-n \bar{J}^a_n.
\end{eqnarray}
Here $\left< JJ \right>'=\eta_{ab} J^a J^b$ and $Q
\eta^{ad}={\ep^a}_{bc} \ep^{dbc}/4=-(1/2 )\eta^{ad}$ is the
quadratic Casimir in the adjoint representation. Writing the nonzero
mode piece as level number $N$, 
$L_0 +\bar{L}_0$ becomes
\begin{eqnarray}
\label{L0'} L_0 +\bar{L}_0&=&N -\left(\f{is}{\hbar} +\f{1}{2}
\right)^{-1} \left< J_0 J_0 \right> ' +\left(\f{is}{\hbar} -\f{1}{2}
\right)^{-1} \left< \bar{J}_0 \bar{J}_0  \right> '
\nonumber \\
&=&N-\left(\f{s^2}{\hbar^2} +\f{1}{4} \right)^{-1} \f{s^2}{4 \hbar^2
{\ell}^2} \left(  \f{4s}{\hbar} r_+ \tilde{r}_{(-)} + r_+^2
-\tilde{r}_{(-)}^2 \right) ,
\end{eqnarray}
where I have used the relations, from (\ref{A}) and (\ref{J}),
\begin{eqnarray}
\left< J_0 J_0-\bar{J}_0 \bar{J}_0\right>' &=&\f{s^2}{\hbar^2}
\f{i
r_+ \tilde{r}_{(-)} }{{\ell}^2}, \nonumber \\
\left< J_0 J_0+\bar{J}_0 \bar{J}_0\right>' &=&\f{s^2}{2 \hbar^2}
\f{r_+^2- \tilde{r}_{(-)}^2 }{{\ell}^2}.
\end{eqnarray}

For {\it non-horizon} boundaries, the Virasoro constraints become
$L_m=0,~\bar{L}_m=0 ~(m \geq 0)$ with
\begin{eqnarray}
L_m \equiv \f{1}{2k +Q} \sum^{\infty}_{n=-\infty} \left<:J_{m-n} J_n
:\right>', ~~\bar{L}_m \equiv  \f{1}{-2k +Q}
\sum^{\infty}_{n=-\infty} \left<:\bar{J}_{m-n} \bar{J}_n :\right>'
\end{eqnarray}
\cite{Fran:97} (for some more general expressions, see Ref.
\cite{Bana:96a}) and ${\cal H} \equiv  L_m -\bar{L}_m,~{\cal
H}_{\varphi} \equiv L_m +\bar{L}_m$ are the diffeomorphism
generators ( in the momentum space) along $t$ and $\varphi$,
respectively. However, for the {\it horizon} boundaries, where the
lapse function--the Lagrange multiplier for ${\cal H}$--vanishes,
${\cal H}$ is not constrained to be zero. Moreover, from the
constancy of the (angular) shift function--the multiplier for ${\cal
H}_{\varphi}$--along $\varphi$, only $L_0 +\bar{L}_0=0$, which does
not depends on $\varphi$, can be implemented \cite{Carl:95,Bana:96}.

From the remnant of the ({\it quantum}) constraint equation $L_0
+\bar{L}_0=0$, which corresponds to the Wheeler-de Witt equation
\cite{Carl:95}, {\it on the horizon} $\partial D_2=r_+$, one obtains
as
\begin{eqnarray}
\label{N}
 N=\f{s^2}{\hbar^2 {\ell}^2} r^2_+ - 4 {\ell}^2 \left( 1 +
\f{\hbar^2}{4 s^2} \right)^{-1} \left(\tilde{r}_{(-)} -\f{2s}{\hbar}
r_+ \right)^2.
\end{eqnarray}
Up to now, the real parameter $\tilde{r}_{(-)}$ is arbitrary and so
we should sum over its possible values to count the number of
macroscopically indistinguishable states. However, in order to
compute the dominant contribution of $N$ to the number of states
$\Omega \approx \mbox{exp}(2\pi \sqrt{c N/6}$)
\cite{Card:86,Park:0402}, it is enough to know its maximum value
$N_{{\rm max}}$, which is attained as
\begin{eqnarray}
\label{N_max}
 N_{{\rm max}}=\f{s^2}{\hbar^2 {\ell}^2}
r^2_+=\left(\f{2 r_{+}}{ \hbar} \right)^2
\end{eqnarray}
for $\tilde{r}_{(-)}=(2s/\hbar) r_+$, in our convention (\ref{s}).
In the semiclassical regime of large $s$, {\it i.e.}, small
$\Lambda$, the central charge may be approximated by that of
six---three $(a=1,2,3)$ for each Kac-Moody sector and there are two
Kac-Moody sectors with the currents $J_n^a,
\bar{J}_n^a$---independent bosonic oscillators such that $c \approx
6$ \cite{Carl:95,Huan:70}. Then the dominant contrbution to the
statistical entropy, through the Cardy's formula
\cite{Card:86,Park:0402,Huan:70} for the asymptotic density of
states of a conformal field theory, is
\begin{eqnarray}
\label{Cardy}
 S_{{\rm st}}=\mbox{ln}~ \Omega \approx 2 \pi \sqrt{\f{c N_{{\rm max}}}{6}}=
 \f{4 \pi r_+}{\hbar}.
\end{eqnarray}
This agrees completely with the thermodynamic entropy (\ref{BH}),
after recovering the Newton's constant $G$. Here note that the
correct $1/\hbar$ factor, whose origin has not been explicit in the
literatures (see \cite{Bana:96} for a comparison), comes from the
definition (\ref{JJ}), essentially. (\ref{Cardy}) is the same result
as in the {\it negative} $\Lambda $, {\it i.e.,} BTZ black hole case
\cite{Bana:96}, which can be recaptured by ${\ell}\rightarrow i
{\ell},~s\rightarrow i s,~\tilde{r}_{(-)} \ra i \tilde{r}_{-}$: The
crucial point is that
(\ref{N}) is {\it invariant} under this substitution, neglecting the
sign change of $(1 -{\hbar^2}/{4 s^2})^{-1}$ which is not important
in the computation of $N_{{\rm max}}$; a similar analysis for the
de-Sitter vacuum $(M=1, J=0)$ has been studied by Strominger and
Maldacena \cite{Mald:98} several years ago in the context of
Carlip's boundary conditions \cite{Carl:95}, but this property has
not been evident in their approach.

To conclude, two remarks are in order.

First, unlike the analyses of the asymptotic isometries
\cite{Park:98}, the Cardy's formula (\ref{Cardy}) is well defined
with the real valued $c$ and $N_{{\rm max}}$. It is essential to
restrict to the horizon in order to acquire the correct BH
entropy.\footnote{~Similar result has been obtained in different
{\it classical} contexts \cite{Carl:99,Park:99PRL,Park:01,Kang:04}.}

Second, note that maximum value of $N$ comes from the $Q$ term
essentially which is purely quantum effect.\footnote{~This seems to
be in contrast to a recent analysis in a different polarization
\cite{Fjel:02}. But the status of the Cardy's formula for this
polarization remains unclear.} Without that term, there is no
$r^2_{+}-\tilde{r}^2_{(-)}$ term in (\ref{L0'}) such that there are
no extremal points of $N$ which satisfies $L_0+\bar{L}_0=0$; rather,
it increases indefinitely as $N=(s/2 \hbar) r_+
\tilde{r}_{(-)}/{\ell}^2$ which has no saddle points so that the
Cardy's formula (\ref{Cardy}) can {\it not} be applied.
This is sharply contrast to the analysis of the asymptotic
isometries of (\ref{A}), where the ``classical'' central charges in
the ( two-copies of ) Virasoro algebra has been crucial to obtain
the correct BH entropy and additional quantum corrections gives only
some small corrections to (\ref{BH}) \cite{Park:01}. Hence, I have
two {\it dual} descriptions of black hole/cosmological horizons. One
is described by the {\it two}-copies of {\it classical} Virasoro
algebra at the infinite boundary and the other's by {\it one}-copy
of the {\it quantum} Virasoro generator through the Sugawara
construction at the horizon $r_+$. By combining the Strominger's
dS/CFT proposal and the horizon holography \cite{Hoof:93} one can
expect that the two descriptions would be closely related. It would
be an outstanding challenge to find a direct connection, or a {\it
RG-flow}, of these two dual descriptions.\footnote{~After writing
this paper, I became aware that Sach and Solodukhin also have
presented a similar idea \cite{Sach:01}.} This will shed some light
on the mystery of dS/CFT correspondence.

\begin{center}
{\bf Acknowledgments}
\end{center}

This work was supported by the Korean Research Foundation Grant
funded by Korea Government(MOEHRD) (KRF-2007-359-C00011).

\end{document}